\documentclass[aps,prl,twocolumn,superscriptaddress,nofootinbib]{revtex4-2}
\usepackage{amsmath,amssymb,amsthm,mathrsfs}
\usepackage{mathtools}
\usepackage{graphicx}
\usepackage{hyperref}
\usepackage{xcolor}
\usepackage{tikz}
\usetikzlibrary{positioning, arrows.meta, shapes.geometric}

\begin{document}

\title{Finite Spectral Recurrences for Loop Integrals in de Sitter and Anti-de Sitter Space}

\author{Bruce A. Bassett}
\affiliation{Wits MIND Institute and School of Computer Science and Applied Mathematics, University of the Witwatersrand, Johannesburg, South Africa}
\affiliation{School for Data Science and Computational Thinking, University of Stellenbosch, South Africa}
\author{Sergio Cacciatori}
\author{Ugo Moschella}
\affiliation{Disat, Universit\`a dell'Insubria, Via Valleggio 11, Como}
\affiliation{INFN, Sez. di Milano, Via Celoria 16, 20133 Milano}

\date{\today}

\begin{abstract}
We introduce a method that does for loop integrals in de Sitter and anti-de Sitter space what integration-by-parts identities do in flat space. A local, coordinate-free propagator algebra common to both geometries yields exact finite-difference equations in spectral labels; with external points they become differential-difference equations, and at the AdS boundary recover known Witten-diagram relations. Acting line by line, they apply in principle to any diagram. We solve them for the two-loop watermelon, isolating its ultraviolet poles, and for one-loop two- and four-point functions.

\end{abstract}

\maketitle
Feynman integrals in de Sitter (dS) and anti-de Sitter (AdS) spaces
are a central challenge in modern theoretical physics, driven by the need to compute inflationary correlators and understand the holographic structure of quantum field theories.
Loop calculations in both spaces  have a long
history, yet explicit evaluations of multiloop graphs remain scarce. A loop integral need not be treated in isolation: in flat space,
integration-by-parts (IBP) identities \cite{Chetyrkin:1981qh,Laporta:2000dsw}
relate it to its neighbours. Curved-space counterparts of this organizing
principle have begun to emerge over the past few years: differential equations for
cosmological correlators \cite{Arkani}, generalized IBP relations for dS
correlators \cite{Chen,ChenFengQinTao}, differential equations in the geodesic
distance for banana diagrams \cite{DiakonovMorozov}, spectral reductions of AdS
loops to tree and contact diagrams \cite{Carmi:2019},
weight-shifting operators in AdS \cite{CostaHansen}, and, closest in spirit to
the present work, difference equations for AdS Mellin amplitudes
\cite{Alaverdian}.

Here we work directly in (Euclidean) spacetime and derive exact finite-difference equations in the spectral labels of the internal lines, which parametrise their bare masses. 
{\em The construction is general and runs identically in dS and AdS, as the propagators share the same contiguity algebra}.

The mechanism is local, and works whether or not the graph has external legs.
For vacuum graphs the relations are algebraic: finite-difference equations in
the spectral labels with polynomial coefficients. They close on a single label
for every banana, at any loop order, and give a finite system of coupled
relations among several labels for more intricate topologies such as the
Mercedes. 

\begin{figure*}[t]
\centering
\resizebox{\textwidth}{!}{%
\begin{tikzpicture}[x=1cm,y=1cm,line width=.35pt,>=Stealth,font=\small]
\node[font=\small] at (1.45,1.55) {2-loop watermelon};
\coordinate (a) at (.65,.35);\coordinate (b) at (2.25,.35);
\draw (a) to[out=65,in=115] (b);\draw (a)--(b);\draw (a) to[out=-65,in=-115] (b);
\fill (a) circle(1.6pt);\fill (b) circle(1.6pt);
\node at (1.45,-.78) {$P_u^\pm(u,v,w)=p_u^\pm(w^2)$};
\node[font=\scriptsize] at (1.45,-1.12) {two loops};
\draw[->] (2.85,.35)--(3.35,.35);

\begin{scope}[xshift=3.2cm]
\node[font=\small] at (1.45,1.55) {1-loop 2-point};
\coordinate (a) at (1,.35);\coordinate (b) at (1.9,.35);
\draw (a) to[out=65,in=115] (b);\draw (a) to[out=-65,in=-115] (b);
\draw (.5,.35)--(a);\draw (b)--(2.4,.35);
\draw[fill=white] (.5,.35) circle(1.4pt);\node[font=\scriptsize,above] at (.5,.35) {$x$};
\draw[fill=white] (2.4,.35) circle(1.4pt);\node[font=\scriptsize,above] at (2.4,.35) {$y$};
\fill (a) circle(1.6pt);\fill (b) circle(1.6pt);
\node at (1.45,-.78) {$p_u^\pm(\mathscr Q_x)$};
\node[font=\scriptsize] at (1.45,-1.12) {one loop};
\end{scope}

\draw[->] (6.05,.35)--(6.55,.35);
\begin{scope}[xshift=6.4cm]
\node[font=\small] at (1.45,1.55) {1-loop 4-point};
\coordinate (a) at (1,.35);\coordinate (b) at (1.9,.35);
\draw (a) to[out=65,in=115] (b);\draw (a) to[out=-65,in=-115] (b);
\foreach \xx/\yy/\lab in {.25/1.03/1,.25/-.33/2} {\draw (\xx,\yy)--(a);\draw[fill=white] (\xx,\yy) circle(1.4pt);\node[font=\scriptsize,left] at (\xx,\yy) {$x_{\lab}$};}
\foreach \xx/\yy/\lab in {2.65/1.03/1,2.65/-.33/2} {\draw (b)--(\xx,\yy);\draw[fill=white] (\xx,\yy) circle(1.4pt);\node[font=\scriptsize,right] at (\xx,\yy) {$y_{\lab}$};}
\fill (a) circle(1.6pt);\fill (b) circle(1.6pt);
\node at (1.45,-.78) {$p_u^\pm(\mathscr Q_{12})$};
\node[font=\scriptsize] at (1.45,-1.12) {one loop};
\end{scope}

\draw[->] (9.34,.35)--(9.85,.35);
\begin{scope}[xshift=9.65cm]
\node[font=\small] at (1.45,1.55) {4-legs Witten graph};
\draw[line width=.35pt] (1.45,.35) ellipse (1.1cm and .8cm);
\coordinate (a) at (1,.35);\coordinate (b) at (1.9,.35);
\draw (a) to[out=65,in=115] (b);\draw (a) to[out=-65,in=-115] (b);
\foreach \xx/\yy/\lab/\pos in {.672/0.958/1/left,.672/-.258/2/left,2.228/.958/3/right,2.228/-.258/4/right} {\coordinate (p\lab) at (\xx,\yy);\draw[fill=white] (p\lab) circle(1.4pt);\node[font=\scriptsize,\pos] at (p\lab) {$P_{\lab}$};}
\draw[densely dashed] (p1)--(a) (p2)--(a) (b)--(p3) (b)--(p4);
\fill (a) circle(1.6pt);\fill (b) circle(1.6pt);
\node at (1.45,-.78) {$p_u^\pm(h^2+\mathcal C_{12})$};
\node[font=\scriptsize] at (1.45,-1.12) {one loop};
\end{scope}
\end{tikzpicture}%
}
\caption{A single local spectral-shift mechanism governs the two-loop watermelon and, after one of its lines is cut, the one-loop two- and four-point graphs with external bulk points and their limit at the AdS boundary, the Witten graphs. Below each graph are the coefficients of its recurrence: numbers for the vacuum graph, polynomials in $\mathscr Q_x$ or in the channel operator for graphs with external points.}
\label{fig:conceptual-spectral-shift}
\end{figure*}
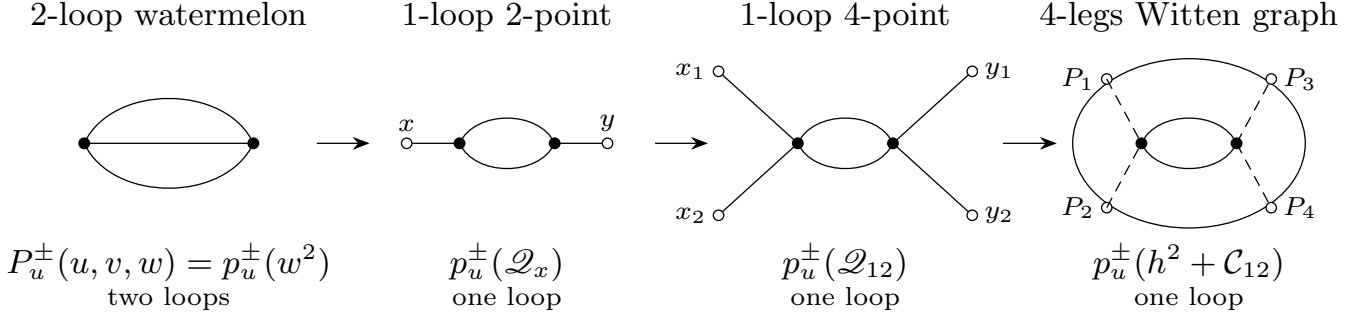

Cutting a line (Fig.~\ref{fig:conceptual-spectral-shift}) turns them into relations for graphs with external
legs: the squared label of the cut line becomes a differential operator
acting on the external coordinates, so the same algebraic relations become
differential-difference equations, difference equations in the internal labels
with coefficients that are differential operators. At the AdS boundary these
operators become conformal Casimirs, and the Witten-diagram difference
equations of Ref.~\cite{Alaverdian} appear as particular cases. The
recurrences thus provide a systematic starting point at higher loop order.

The method is also constructive. We solve explicitly the finite-difference
equations for the watermelons and the one-loop two-point functions, and reduce
the one-loop four-point functions to tree-level diagrams, in both dS and AdS.
For the watermelons we obtain new representations in which all
ultraviolet divergences are isolated in explicit Gamma factors, while the
remaining hypergeometric series are regular near $d=4$. This yields
the four-dimensional ultraviolet poles in closed form and makes the constant
term explicit. Together, these results make the renormalized two-loop
effective potential in four-dimensional dS and AdS space, exact in curvature and masses, accessible for the first time, bringing it to the level reached
long ago in flat space in Ref.~\cite{Ford:1992pn}.

\emph{A local mass ladder.}---We work at unit radius on the $d$-dimensional sphere \(S^d\)   and Euclidean AdS  
\(H^d\), distinguished by \(\sigma=+1\) and \(\sigma=-1\), respectively. Put
$h=a+1=(d-1)/2$, $\mu_q=h^2-q^2$ and $\kappa_q=a^2-q^2$; the mass
squared is $\sigma\mu_q$. The invariant variable is $z=X\cdot Y=\cos r$ on
$S^d$ and $z=X\cdot Y=\cosh r$ on $H^d$.
The relevant propagators are Schwinger functions of maximally analytic two-point functions 
\cite{Cacciatori:2024ds,Cacciatori:2024ads} 
\begin{equation}\label{eq:props}
\begin{aligned}
  G^+_q&=\frac{\Gamma(h+q)\Gamma(h-q)}{(4\pi)^{h+\frac 1 2}\Gamma(h+\frac 12)}\,
  {}_2F_1\Big(h+q,h-q;h+\tfrac 12;\tfrac{1+z}{2}\Big)\\[2pt]
  G^-_q&=\frac{\Gamma(h+q)}{2\pi^{h}\Gamma(q+1)}\,(2z)^{-h-q}\,
  {}_2F_1\Big(\tfrac{h+q}{2},\tfrac{h+q+1}{2};q+1;\tfrac1{z^2}\Big) 
\end{aligned}
\end{equation}
 and solve the equation 
\begin{equation}\label{eq:green}
  (-\nabla^2+\sigma\mu_q)\,G_q^\sigma(X,Y)=\delta(X,Y).
\end{equation}
They are meromorphic in $q$ and enjoy the shift identities
\begin{equation}\label{eq:ladder}
\begin{aligned}
  (q-h+1)G^\sigma_{q+1}&=(q+h)z\,G^\sigma_q-(1-z^2)\,{G_q^\sigma}',\\
  (q+h-1)G^\sigma_{q-1}&=(q-h)z\,G^\sigma_q+(1-z^2)\,{G^\sigma_q}' ,
\end{aligned}
\end{equation}
which turn multiplication by $z$ and differentiation into integer displacements of
$q$. This is the first ingredient, and it is the same in both geometries.
The shifts move the labels off the physical values.  For a dS principal-series
field $q=i\nu$ is imaginary, so $q\pm1$ has complex mass. On AdS, repeated
shifts violate $q>-1$, the unitarity bound. The recurrence requires them all. The identities derived below are therefore identities of the amplitude as a
meromorphic function of the labels, exactly as flat-space IBP relates physical
integrals to an analytic family with shifted propagator powers. 

The tadpole shows the mechanism at its simplest. It is the 
regularized propagator at coincidence $T^\sigma_q=G^\sigma_q(1)$
\begin{equation}\label{eq:tad} 
T^\sigma_q=\frac{\Gamma(1-h-\frac12)}{(4\pi)^{h+\frac12}} \times\begin{cases} \dfrac{\Gamma(h+q)\Gamma(h-q)}{\Gamma(\frac12+q)\Gamma(\frac12-q)}, & \sigma=+1,\\[2ex] \dfrac{\Gamma(h+q)}{\Gamma(1-h+q)}, & \sigma=-1 . \end{cases}
\end{equation}
Setting $z=1$ in \eqref{eq:props} and \eqref{eq:ladder} is legitimate for $\Re d<2$ and leads to
a homogeneous two-term recurrence
\begin{equation}\label{eq:tadrec}
  (q-h)(q-h+1)\,T^\sigma_{q+1}=(q+h)(q+h-1)\,T^\sigma_{q-1},
\end{equation}
identical on $S^d$ and $H^d$ 
\footnote{On $H^d$, Eq.~\eqref{eq:tadrec} is the tadpole difference equation of Ref.~\cite{Alaverdian} under the identification $\delta=q$ of their spectral variable with our mass label.}
although the tadpoles themselves are not.

 The ladder
\eqref{eq:ladder} is local and moves masses, but to relate
different graphs we need more: Green's identity, the second ingredient, whose
surviving terms are always contact terms. The two-line banana (bubble) 
\begin{equation}
  \mathcal{I}_2^\sigma(q_1,q_2)=\int dY\;G^\sigma_{q_1}(X,Y)\,G^\sigma_{q_2}(X,Y),
\end{equation}
shows this in its purest form. The integrand
of
\begin{equation}\label{eq:green2}
  \int dY\,\big(G^\sigma_{q_1}\nabla^2G^\sigma_{q_2}-G^\sigma_{q_2}\nabla^2G^\sigma_{q_1}\big)=0
\end{equation}
is a divergence; the surface term vanishes on $S^d$, while $\Re(q_1+q_2)>0$
suffices on $H^d$.  On inserting
\eqref{eq:green} 
we get  
\begin{equation}\label{eq:bub}
  \sigma\,\mathcal{I}_2^\sigma(q_1,q_2)=\frac{T^\sigma_{q_1}-T^\sigma_{q_2}}{q_1^2-q_2^2} .
\end{equation}
The bubble is thus not merely reduced but evaluated,   reproducing in a single step the result of Refs.~\cite{Cacciatori:2024ds,Cacciatori:2024ads}.

\emph{The watermelon.}---With the above two ingredients in hand we can fully
solve the first nontrivial example, the three-line banana
$\mathcal{I}_3^\sigma(\mathbf q)=\int dY\,G^\sigma_{q_1}G^\sigma_{q_2}G^\sigma_{q_3}$,
with $\mathbf q=(q_1,q_2,q_3)$, unit shifts $\mathbf e_k$, and
$\{i,j,k\}=\{1,2,3\}$ throughout. Green's identity is bilinear, so a third line
must be carried outside the Wronskian by something whose derivative returns a
propagator: this is the primitive $\partial_zQ^\sigma_q=-G^\sigma_q$, itself a
shift of the label,
\begin{equation}\label{eq:prim}
 Q^\sigma_q=\frac{G^\sigma_{q-1}-G^\sigma_{q+1}}{2q}.
\end{equation}
The current is then fixed as
\begin{equation}\label{eq:current}
 \mathcal{C}_{ij|k}=[\sigma(1-z^2)]^{h+\frac12}\,\big(G^\sigma_{q_j}{G^\sigma_{q_i}}'
 -G^\sigma_{q_i}{G^\sigma_{q_j}}'\big)\,Q^\sigma_{q_k}.
\end{equation}
Integrating
$\partial_z\,\mathcal{C}_{ij|k}$ gives, from its
three orientations, two
independent relations
\begin{equation}\label{eq:sourced}
 \big[\kappa_i\mathsf D_i-\kappa_j\mathsf D_j+(\kappa_i-\kappa_j)\mathsf D_k\big]
 \mathcal{I}_3^\sigma=\frac{2h-1}{\sigma\kappa_k}\big(T^\sigma_j-T^\sigma_i\big)T^\sigma_k,
\end{equation}
where the operator $\mathsf D_m$ replaces $G^\sigma_{q_m}$ with $Q^\sigma_{q_m}$.
Two more relations come from the identities ($k=1,2,3$)
\begin{equation}\label{eq:aux}
 \mathcal{J}^\sigma=\frac{(q_k+a)\,\mathcal{I}_3^\sigma(\mathbf q-\mathbf e_k)
 +(q_k-a)\,\mathcal{I}_3^\sigma(\mathbf q+\mathbf e_k)}{2q_k},
\end{equation}
where $\mathcal{J}^\sigma=\int dY\,z\,G^\sigma_{q_1}G^\sigma_{q_2}G^\sigma_{q_3}$,
and a fifth is
\begin{equation}\label{eq:fifth}
 4(h-1)\,\mathcal{J}^\sigma=\sum_{m=1}^{3}\kappa_m\mathsf D_m\mathcal{I}_3^\sigma.
\end{equation}
These five linear relations among the six neighbours
$\mathcal I_3^\sigma(\mathbf q\pm\mathbf e_m)$ leave one direction undetermined:
the vector $\mathbf P=(P_1^+,P_1^-,\dots,P_3^-)$ that satisfies all five with
vanishing source, where
\[
 P_k^{\pm}=\prod_{\epsilon,\epsilon'=\pm1}(a\pm q_k+\epsilon q_i+\epsilon'q_j)
 =\lambda-2a(a\pm q_k)\,\frac{\partial\lambda}{\partial\kappa_k},
\]
with $\lambda=\kappa_1^2+\kappa_2^2+\kappa_3^2-2(\kappa_1\kappa_2+\kappa_2\kappa_3+\kappa_3\kappa_1)$
the K\"all\'en function. For generic dimension and spectral labels,
$\mathbf P$ spans the kernel. A linear combination of the six neighbouring
integrals is thus determined by the five relations precisely when its coefficient
vector is orthogonal to $\mathbf P$. For the two neighbours of
a single label $k$, this fixes the ratio of the coefficients and gives the difference equation
\begin{align}
 &P_k^-\,\mathcal{I}_3^\sigma(\mathbf q+\mathbf e_k)
 -P_k^+\,\mathcal{I}_3^\sigma(\mathbf q-\mathbf e_k)
 =2\sigma q_k(2h-1)\nonumber\\
 &\quad\times\Big[2T_iT_j-(T_i+T_j)T_k
 +\frac{\kappa_i-\kappa_j}{\kappa_k}(T_i-T_j)T_k\Big],
 \label{eq:rec}
\end{align}
with $T_m=T^\sigma_{q_m}$. Remarkably,  Eq. \eqref{eq:rec} can be solved in closed form. The general solution of the homogeneous recurrence  can be
written as $H(q_k)\,p(q_k)$, where
\begin{equation}
H(q_k)=\prod_{\boldsymbol\epsilon}\frac{\Gamma(B_{\boldsymbol\epsilon})}
 {\Gamma(B_{\boldsymbol\epsilon}-a)},\ \ B_{\boldsymbol\epsilon}=\tfrac12(h+q_k+\epsilon_1q_i+\epsilon_2q_j), \nonumber
\end{equation}
and $p$ any function of period two, 
hence constant along the lattice $q_k\in q_0+2\mathbb N$.  Telescoping gives a
particular solution of the inhomogeneous recurrence. 
The physical value of $p$ that gives the correct ${\cal I}_3$ can be fixed by examining the solution along the lattice at large $|m_k^2|$, where
$m_k^2=\sigma(h^2-q_k^2)$ is the squared mass of the line:
heavy on $H^d$ and tachyonic on $S^d$. In both cases the integral 
 approaches the flat-space sunset with two massless lines, $b_h\,(m_k^2)^{d-3}$---on $S^d$
continued to negative mass squared, with the standing wave fixed by the compact
geometry---where
\begin{equation}
 b_h=\frac{\Gamma(h-\frac12)^2\,\Gamma(\frac32-h)\,\Gamma(2-2h)}
 {(4\pi)^{2h+1}\,\Gamma(h+\frac12)}.
\end{equation}
Hence $\mathcal I_3^\sigma/H\to2^{4a}b_h\,s_k^\sigma$, with
$s_k^-=1$ and the sphere's standing-wave factor
$s_k^+=-\sin\pi(h+q_k)/\sin\pi(h-q_k)$. This fixes $p$, and the solution is
\begin{align}
 \mathcal{I}_3^\sigma={}&2^{4a}b_h\,s_k^\sigma\,H(q_k)
 -\frac{\sigma(q_k+1)(2h-1)}{8\prod_{\boldsymbol\epsilon}B_{\boldsymbol\epsilon}}
 \Big[2T_iT_j\mathcal F_1\nonumber\\
 &-(T_i+T_j)\bar T_k\mathcal F_2
 +\frac{\kappa_i-\kappa_j}{\bar\kappa_k}(T_i-T_j)\bar T_k\,\mathcal F_3\Big],
 \label{wm}
\end{align}
where $\bar T_k=T^\sigma_{q_k+1}$, $\bar\kappa_k=\kappa_{q_k+1}$, the second term
decays relative to $H$ along the lattice for $\Re d>2$ (elsewhere the result
follows by continuation), and, at unit argument,
\begin{align}
 \mathcal F_1&={}_6F_5\Big[{B_{\boldsymbol\epsilon}-a,\,\frac{q_k+3}{2},\,1
 \atop B_{\boldsymbol\epsilon}+1,\,\frac{q_k+1}{2}}\Big],\nonumber\\
 \mathcal F_2&={}_8F_7\Big[{B_{\boldsymbol\epsilon}-a,\,\frac{q_k+3}{2},\,
 \frac{q_k+2+a}{2},\,\frac{q_k+3+a}{2},\,1
 \atop B_{\boldsymbol\epsilon}+1,\,\frac{q_k+1}{2},\,\frac{q_k+1-a}{2},\,
 \frac{q_k+2-a}{2}}\Big],\nonumber\\
 \mathcal F_3&={}_8F_7\Big[{B_{\boldsymbol\epsilon}-a,\,\frac{q_k+3}{2},\,
 \frac{q_k+1+a}{2},\,\frac{q_k+2+a}{2},\,1
 \atop B_{\boldsymbol\epsilon}+1,\,\frac{q_k+1}{2},\,\frac{q_k+2-a}{2},\,
 \frac{q_k+3-a}{2}}\Big].\nonumber
\end{align}

The dS and AdS watermelons were first evaluated by two of us in Refs. \cite{Cacciatori:2024ds,Cacciatori:2024ads} using the
K\"all\'en--Lehmann representation of the product of two propagators \cite{bros1,bros2}; in $d=4$ we found a closed formula for the double pole using the Erd\'elyi--Tricomi theorem and outlined a finite-part procedure that however proved exceedingly difficult to carry to completion. The new formula
(\ref{wm}), numerically identical to our previous results but not related to
them by any transformation we are aware of, removes this obstacle: all  
ultraviolet poles sit in explicit Gamma factors, while the
remaining hypergeometric series are regular near $d=4$. 
In particular, for dS at $d=4-2\varepsilon$, writing
\[
\mathcal{I}_3^+=\frac{1}{256\pi^4}
\left[C_{-2}\varepsilon^{-2}+C_{-1}\varepsilon^{-1}+C_0\right]+O(\varepsilon),
\]
with the labels $q_i$ held fixed, we obtain
\begin{align}
& C_{-2}=\frac{q_1^2+q_2^2+q_3^2}{2}-\frac38,\label{eq:C2}\cr
& C_{-1}=\frac{17}{8}+\frac{q_1^2+q_2^2+q_3^2}{2} -\big(\log4\pi+1-\gamma_E\big)\Big(\frac34-q_1^2\nonumber\\&-q_2^2-q_3^2\Big)
+\sum_{j=1}^{3}\Big(\frac14-q_j^2\Big)
 \Big[\psi\Big(\frac12+q_j\Big)+\psi\Big(\frac12-q_j\Big)\Big].
\nonumber
\end{align}
These coefficients, together with the explicit representation of the finite
part $C_0$ (not reproduced here), provide the essential ingredients for the
renormalized two-loop effective potential of the $O(N)$ model
\cite{companion}, a calculation that had remained
inaccessible to all previous methods in $d=4$ (for $d=3$ see Ref.~\cite{bonanno}).

\emph{Arbitrary finite-loop bananas and multivertex topologies.}---The same mechanism yields a finite scalar recurrence for bananas with an arbitrary number of lines, although the associated current no longer factorizes. For \(M\) lines, we expand it over the \(2^M\) patterns
$
\prod_{i\in S}(G^\sigma_{q_i})'
\prod_{j\notin S}G^\sigma_{q_j},
$
with unknown polynomial coefficients. Requiring that only the two patterns associated with the shifted line survive gives a finite linear system for these coefficients.
For the next simplest case, the four-line basketball, the sixteen unknown
polynomials contain forty coefficients. Eliminating them yields a recurrence
involving a single spectral label, with five terms separated by steps of two:
\begin{equation}\label{eq:basket}
  \sum_{r=-2}^{2}\ell_{2r}(h,\mathbf{q})\,  \mathcal{I}_{4}^\sigma(\mathbf{q}+2r\mathbf{e}_k)  = \sigma \sum_{i=1}^{4}s_i(h,\mathbf{q})  \prod_{j\ne i}T_{q_j}^\sigma .\end{equation}
The coefficients are generically nonzero polynomial determinants, which we do
not reproduce here, and the source is cubic in tadpoles. At equal masses the
recurrence simplifies  and only two polynomials are needed.
Define the operator acting on the fourth line by
$\mathsf Z_q f(q)=\frac{q-a}{2q}f(q+1)+\frac{q+a}{2q}f(q-1)$, powers denoting
composition, with the coefficients shifted at each step. Then
\[
\Bigl[(\alpha-\beta)\,\mathsf Z_q+\beta\,\mathsf Z_q^3\Bigr]
\mathcal{I}_{4}^{\sigma}(q,q,q,q)=24\sigma(2h-1)^3(T_q^\sigma)^3,
\]
where
\[
\alpha(q)=(2h-2)(6h-5)\bigl[12q^2-8h^3+12h^2+8hq^2-6\bigr],
\]
\[
\beta(q)=(2h-3)\bigl[4q^2-(2h-3)^2\bigr]\bigl[16q^2-(2h-3)^2\bigr].
\]
The relation involves the four graphs whose fourth label is displaced by
$\pm1$ and $\pm3$, the other three held at $q$, and its source is simply the
cube of a tadpole.

For the six-line Mercedes graph, the construction instead produces a finite system of coupled mass-shift equations: each equation contains two terms in a selected spectral label, with operator-valued coefficients that shift the other five. The contact source is a specific contracted five-edge graph and therefore retains the three-loop structure, rather than factorizing into a product of tadpoles. Details of the general construction will be presented elsewhere.

\emph{Cutting a line: external bulk points.}---Difference equations are not
restricted to vacuum diagrams. Cutting one line of the watermelon leaves the
kernel $K^\sigma_{u,v}=G^\sigma_uG^\sigma_v$. The removed label $w$ enters the
coefficients of Eq.~\eqref{eq:rec} only through   the eigenvalue $w^2$
of $\mathscr Q=h^2-\sigma\nabla^2$ on the propagator (up to a contact term).
Writing
$P_u^\pm(u,v,w)=p_u^\pm(w^2)$ and applying the ladder to $K^\sigma_{u,v}$, we
obtain the same difference equation with $w^2$ replaced by $\mathscr Q$,
together with a source localized at coincidence:
\begin{equation}\label{eq:opened-kernel}
 p_u^-(\mathscr Q)K^\sigma_{u+1,v}
 -p_u^+(\mathscr Q)K^\sigma_{u-1,v}
 =2\sigma u(2h-1)\,S^\sigma_u(\mathscr Q)\,\delta ,
\end{equation}
with $S^\sigma_u(t)=2T^\sigma_v+(T^\sigma_u/\kappa_u)(u^2+v^2-a^2-t)$.

Attaching one external propagator at each vertex (Fig.~\ref{fig:conceptual-spectral-shift}) gives the one-loop two-point 
function
\[
 C^\sigma_{u,v}(x,y)=\int dX\,dY\,G^\sigma_r(x,X)K^\sigma_{u,v}(X,Y)G^\sigma_s(Y,y).
\]
Integrating \eqref{eq:opened-kernel} against the legs, $\mathscr Q$ moves onto
$x$, the contact collapses one vertex, and the resolvent identity
$G^\sigma_r\circ G^\sigma_s=\sigma(G^\sigma_r-G^\sigma_s)/(r^2-s^2)$ cancels
the factor $\sigma$ giving rise to the difference-differential equation for the one-loop two-point function
\begin{align}
&p_u^-(\mathscr Q_x)C^\sigma_{u+1,v}-p_u^+(\mathscr Q_x)C^\sigma_{u-1,v}\nonumber\\
&\quad=\frac{2u(2h-1)}{r^2-s^2}\big[S^\sigma_u(r^2)G^\sigma_r-S^\sigma_u(s^2)G^\sigma_s\big].
\label{open}
\end{align}
Attaching instead two legs at each vertex,
$f^\sigma(X)=G^\sigma_{r_1}(x_1,X)G^\sigma_{r_2}(x_2,X)$ and
$g^\sigma(Y)=G^\sigma_{s_1}(Y,y_1)G^\sigma_{s_2}(Y,y_2)$, gives the four-point
function $F^\sigma_{u,v}=\int dX\,dY\,f^\sigma K^\sigma_{u,v}\,g^\sigma$. Now
$\mathscr Q$ cannot be moved onto a single point. But $f^\sigma$ is invariant
under simultaneous isometries of its three arguments, so the Laplacian at the
vertex equals the quadratic Casimir $\mathcal C_{12}$ of the isometry group
acting on the pair $(x_1,x_2)$, $\nabla^2_Xf^\sigma=\mathcal C_{12}f^\sigma$;
its eigenvalues label the representations exchanged in the channel $(12)$.
With $\mathscr Q_{12}=h^2-\sigma\mathcal C_{12}$, the difference-differential equation for the one-loop four-point function
\begin{align}
&p_u^-(\mathscr Q_{12})F^\sigma_{u+1,v}-p_u^+(\mathscr Q_{12})F^\sigma_{u-1,v}\nonumber\\
&\quad=2\sigma u(2h-1)\,S^\sigma_u(\mathscr Q_{12})\,D^\sigma,
\label{eq:four-bulk}
\end{align}
where $D^\sigma=\int dX\,f^\sigma g^\sigma$ is the tree contact diagram. Here
$\sigma$ survives, since no resolvent evaluates the contact.

Both equations are solved at once. Convolution with $K^\sigma_{u,v}$ commutes
with $\mathscr Q$ and acts at spectral value $t$ as a multiplier
$\widehat K^\sigma_{u,v}(t)$, which solves the watermelon recurrence with
$w^2\to t$ and source the coefficient of $T^\sigma_w$. The solution is therefore the
coefficient of $T^\sigma_w$ in Eq.~\eqref{wm}, with $k=u$ and $(i,j)=(v,w)$:
\begin{align}
\widehat K^\sigma_{u,v}(w^2)={}&-\frac{\sigma(u+1)(2h-1)}{8\prod_{\boldsymbol\epsilon}B_{\boldsymbol\epsilon}}
\Big[2T^\sigma_v\mathcal F_1-\bar T_u\mathcal F_2\nonumber\\
&-\frac{\kappa_v-\kappa_w}{\bar\kappa_u}\,\bar T_u\,\mathcal F_3\Big].
\label{eq:Khat}
\end{align}
Since $C^\sigma_{u,v}$ and $F^\sigma_{u,v}$ are linear in the kernel, inverting
the spectral transform gives both. On the sphere the spectrum is discrete,
$t_\ell=(\ell+h)^2$, with eigenfunctions
$P_\ell(z)={}_2F_1\big(-\ell,\ell+2h;\tfrac d2;\tfrac{1-z}{2}\big)$, and with
$z=x\cdot y$
\begin{align}
 C^+_{u,v}&=\sum_{\ell=0}^{\infty}\frac{{(2\ell+d-1)\Gamma(\ell+d-1)\Gamma(\frac{d+1}{2})}
\,\widehat K^+_{u,v}(t_\ell)\,
 P_\ell(z)}{{2\pi^{(d+1)/2}\,\Gamma(d)\,\ell!}(t_\ell-r^2)(t_\ell-s^2)},\nonumber\\
 F^+_{u,v}&=\sum_{\ell=0}^{\infty}\widehat K^+_{u,v}(t_\ell)\,E^{(\ell)},
 \label{eq:opened-sphere-expansion}
\end{align}
$
 E^{(\ell)}=\int dX\,dY\,f^+(X)\,\mathcal N_\ell\,P_\ell(X\cdot Y)\,g^+(Y)
$
is the tree diagram with the internal propagator replaced by the projector
onto the harmonics of degree $\ell$. On $H^d$ the inversion can be deformed
onto the poles of $\widehat K^-_{u,v}$, which lie at the two-particle
thresholds $t=w_n^2$, $w_n=h+u+v+2n$. Each pole becomes a free propagator, so
that, with $\rho_n=-\mathrm{Res}_{t=w_n^2}\widehat K^-_{u,v}$,
\begin{align}
 &K^-_{u,v}=\sum_{n\ge0}\rho_n\,G^-_{w_n},\nonumber\\
 &C^-_{u,v}=\sum_{n\ge0}\rho_n\,G^-[r^2,s^2,w_n^2],\quad
 F^-_{u,v}=\sum_{n\ge0}\rho_n\,E^-_{w_n}.
 \label{eq:opened-hyperbolic-expansion}
\end{align}
Here $G^-[r^2,s^2,w^2]=G^-_r\circ G^-_w\circ G^-_s$ is the second divided
difference of $G^-_q$ in $q^2$, and $E^-_w=\int dX\,dY\,f^-G^-_wg^-$ is the
tree exchange diagram. The first relation is the K\"all\'en--Lehmann expansion
of the kernel, with weights $\rho_n$ that coincide with those of
Ref.~\cite{bros2}, and the one-loop bubble is thus a sum of tree exchanges at
the thresholds, as implied by that expansion. Here it follows from the
recurrence rather than being assumed, and the same construction applies on
the sphere.

\emph{AdS boundary points: Witten diagrams.}---When the external points of the
four-point graph are taken to the boundary of $H^d$, the propagators attached
to the vertices become bulk-to-boundary propagators $\mathcal E_i$ of
dimensions $\Delta_i$ at boundary points $P_i$. With
$f=\mathcal E_1\mathcal E_2$ and $g=\mathcal E_3\mathcal E_4$, the one-loop
Witten diagram and the tree contact diagram are
\begin{equation}\label{eq:witten-definition}
\begin{aligned}
 W_{u,v}&=\int dX\,dY\,f(X)\,K^-_{u,v}(X,Y)\,g(Y),
\end{aligned}
\end{equation}
The AdS isometries act on the boundary points as conformal transformations, so
the Casimir $\mathcal C_{12}$ becomes the conformal Casimir of the channel,
with eigenvalue $c_{\delta,\ell}=\delta(\delta-2h)+\ell(\ell+2h-2)$ on an
exchanged operator of dimension $\delta$ and spin $\ell$, and
Eq.~\eqref{eq:four-bulk} becomes, with $\mathscr Q_{12}=h^2+\mathcal C_{12}$,
\begin{equation}\label{eq:four}
\begin{aligned}
 &p_u^-(\mathscr Q_{12})\,W_{u+1,v}-p_u^+(\mathscr Q_{12})\,W_{u-1,v}\\
 &\qquad=-2u(2h-1)\,S_u^-(\mathscr Q_{12})\,D ,
\end{aligned}
\end{equation}
whose source combines the contact diagram $D=\int dX\,f(X)\,g(X)$ and the Casimir acting on it, with
tadpole coefficients. After matching normalizations, this relation reproduces
the four-point difference equation of Ref.~\cite{Alaverdian} for the one-loop
bubble. Projected onto a single conformal partial wave it gives their
recurrence in the internal dimensions, and a recurrence for the regular radial
eigenfunction gives their recurrence in the external dimensions, both for the
coefficient that remains once the regulated boundary two-point factor is
removed. As throughout, these identities hold in a common convergence domain
and extend by meromorphic continuation; a renormalized version must include
the counterterms of the chosen scheme.

\emph{Conclusions.}---We have introduced a method that does for loop integrals
in de Sitter and anti-de Sitter space what integration-by-parts identities do
in flat space: it organizes them into families related by exact equations,
finite-difference equations in the spectral labels of the lines for vacuum
graphs, and differential-difference equations once external points are
retained.

Like IBP
identities, its ingredients are local and independent of the particular graph:
propagator contiguity shifts the labels, and Green's identity produces contact
terms. Both are shared by the two geometries, require no choice of coordinates,
and act line by line. The same algebra governs vacuum graphs, graphs with
external bulk points, where the spectral labels become differential operators,
and Witten diagrams, where they become conformal Casimirs. The equations,
sources included, have the same form in both geometries: the geometry enters
only through the propagators and tadpoles of each space and through global
data, the surface terms and the condition that selects the physical solution.

The method is constructive as well. The two-loop watermelon is solved in closed
form in both geometries, with its ultraviolet structure explicit; the one-loop
two-point function follows in position space, the one-loop four-point function
reduces to tree diagrams, and the AdS K\"all\'en--Lehmann expansion is
recovered from the recurrence alone.

In flat space, IBP identities turned multiloop calculations from case-by-case
integration into a systematic algebraic problem, and made possible the
precision computations on which perturbative quantum field theory now rests.
We expect the finite-difference equations introduced here to play the same role
in curved space, and quantum field theory in de Sitter and anti-de Sitter space
to reach, with this tool, the level of analytic control long available in flat
space.

\emph{AI-assisted research \& verification.}---This research utilised approximately 120 hours of multi-agent AI compute primarily across GPT-6 Astra Pro and  Claude Fable 5.1,
with up to 23 subagents utilised in any computation. The key ideas and writing are human. 
Responsibility for the final manuscript remains with the authors.

\begin{acknowledgments}
We thank the Institut d'\'Etudes Scientifiques de Carg\`ese, where this work
was started during the workshop \emph{Hot Topics in Modern Cosmology},
Spontaneous Workshop XVIII.
\end{acknowledgments}


\begin{thebibliography}{99}

\bibitem{Chetyrkin:1981qh}
K.~G.~Chetyrkin and F.~V.~Tkachov,
Nucl.\ Phys.\ B \textbf{192}, 159 (1981).

\bibitem{Laporta:2000dsw}
S.~Laporta,
Int.\ J.\ Mod.\ Phys.\ A \textbf{15}, 5087 (2000),
\href{https://arxiv.org/abs/hep-ph/0102033}{arXiv:hep-ph/0102033}.

\bibitem{Arkani}
N.~Arkani-Hamed, D.~Baumann, A.~Hillman, A.~Joyce, H.~Lee and G.~L.~Pimentel,
J.\ High Energy Phys.\ \textbf{09} (2025) 009,
\href{https://arxiv.org/abs/2312.05303}{arXiv:2312.05303}.

\bibitem{Chen}
J.~Chen and B.~Feng,
J.\ High Energy Phys.\ \textbf{06} (2024) 199,
\href{https://arxiv.org/abs/2401.00129}{arXiv:2401.00129}.

\bibitem{ChenFengQinTao}
J.~Chen, B.~Feng, Z.~Qin and Y.~X.~Tao,
\href{https://arxiv.org/abs/2604.14549}{arXiv:2604.14549} (2026).

\bibitem{DiakonovMorozov}
D. Diakonov and A. Morozov,
Phys.\ Lett.\ B \textbf{858}, 139079 (2024),
\href{https://arxiv.org/abs/2408.15724}{arXiv:2408.15724}.

\bibitem{Carmi:2019}
D.~Carmi,
J.\ High Energy Phys.\ \textbf{06} (2020) 049,
\href{https://arxiv.org/abs/1910.14340}{arXiv:1910.14340}.
J.\ High Energy Phys.\ \textbf{07} (2021) 186,
\href{https://arxiv.org/abs/2104.10500}{arXiv:2104.10500}.
J.\ High Energy Phys.\ \textbf{08} (2024) 193,
\href{https://arxiv.org/abs/2402.02481}{arXiv:2402.02481}.

\bibitem{CostaHansen}
M.~S.~Costa and T.~Hansen,
J.\ High Energy Phys.\ \textbf{09} (2018) 040,
\href{https://arxiv.org/abs/1805.01492}{arXiv:1805.01492}.

\bibitem{Alaverdian}
M.~Alaverdian, A.~Herderschee, R.~Roiban and F.~Teng,
J.\ High Energy Phys.\ \textbf{12} (2024) 070,
\href{https://arxiv.org/abs/2406.04186}{arXiv:2406.04186}.

\bibitem{Ford:1992pn}
C.~Ford, I.~Jack and D.~R.~T.~Jones,
Nucl.\ Phys.\ B \textbf{387}, 373 (1992);
\textbf{504}, 551(E) (1997),
\href{https://arxiv.org/abs/hep-ph/0111190}{arXiv:hep-ph/0111190}.

\bibitem{Cacciatori:2024ds}
S.~L.~Cacciatori, H.~Epstein and U.~Moschella,
J.\ High Energy Phys.\ \textbf{07} (2024) 182,
\href{https://arxiv.org/abs/2403.13145}{arXiv:2403.13145}.

\bibitem{Cacciatori:2024ads}
S.~L.~Cacciatori, H.~Epstein and U.~Moschella,
J.\ High Energy Phys.\ \textbf{08} (2024) 109,
\href{https://arxiv.org/abs/2403.13142}{arXiv:2403.13142}.

\bibitem{bros1}
J.~Bros, H.~Epstein, M.~Gaudin, U.~Moschella and V.~Pasquier,
Commun.\ Math.\ Phys.\ \textbf{295}, 261 (2010),
\href{https://arxiv.org/abs/0901.4223}{arXiv:0901.4223}.

\bibitem{bros2}
J.~Bros, H.~Epstein, M.~Gaudin, U.~Moschella and V.~Pasquier,
Commun.\ Math.\ Phys.\ \textbf{309}, 255 (2012),
\href{https://arxiv.org/abs/1107.5161}{arXiv:1107.5161}.

\bibitem{companion} In preparation.

\bibitem{bonanno}
A.~Bonanno, S.~L.~Cacciatori and U.~Moschella,
Symmetry \textbf{18}, 801 (2026),
\href{https://arxiv.org/abs/2603.02140}{arXiv:2603.02140}.
\end{thebibliography}
\end{document}